\documentclass[
 prab, 
 twocolumn, 
 showkeys, 
 floatfix, 
 amsmath,
 amssymb,
 aps
]{revtex4-2}
\usepackage{svg}
\usepackage{graphicx}
\usepackage{dcolumn} 
\usepackage{bm}
\usepackage{comment}
\usepackage{multirow}
\usepackage{fancyvrb}
 
\usepackage{listings}

\lstdefinelanguage{pseudocode}{
  morekeywords={end,for,each,if,then,else,return,from,to},
  sensitive=false,
  morecomment=[l]{//}
}
\usepackage{hyperref}
\begin{document}

\preprint{APS/123-QED}

\title{\textbf{A modern particle-in-cell code for beam-beam simulations of future colliders}}

\author{Arianna Formenti, Remi Lehe, Axel Huebl, Stepan S. Bulanov, Jens Osterhoff, Andrew Myers, Weiqun Zhang, Jean-Luc Vay}
\affiliation{
Accelerator Technology and Applied Physics Division, Lawrence Berkeley National Laboratory, Berkeley, California, USA}
\email{Contact author: ariannaformenti@lbl.gov}

\author{Spencer Gessner, Peter Kicsiny, Anna Kinderman, Bao Nguyen, Jinseo Park}
\affiliation{SLAC, Menlo Park, California, USA}

\date{\today}

\begin{abstract}
A key challenge in the design of next-generation colliders is predicting the effects of beam-beam interactions and using these predictions to optimize machine performance. Beam-beam interactions are characterized by strong focusing forces, radiative effects, and the generation of background particles. While analytical formulas provide insight into the dynamics of beam-beam interactions, these nonlinear, coupled effects are most accurately described with numerical methods. In this paper, we introduce the open-source particle-in-cell code WarpX as a modern tool for modeling the interaction point in future lepton colliders. WarpX provides high-performance execution on distributed GPU architectures within an open-source and actively maintained code base. We present convergence scans and benchmarks against the established beam-beam codes GUINEA-PIG (v1.2.2) and CAIN (v2.42) for luminosity spectra, photon emission, and secondary-particle production under a wide range of beam parameters corresponding to different collider concepts. Our results show that WarpX (v26.09) reproduces the benchmark beam-beam observables while enabling GPU-portable, extensible simulations. These results establish WarpX as a scalable platform for high-fidelity beam-beam modeling in future collider studies.
\end{abstract}


\maketitle

\section{\label{sec:introduction}Introduction}
Future colliders must operate with extremely high luminosities to study rare processes with small cross sections~\cite{Shiltsev2021RevModPhys, Roser2023, Arduini:2025hor}. 
High luminosities are achieved by colliding dense bunches at the interaction points (IPs).
Here, the electromagnetic (EM) fields of the colliding bunches induce kicks on individual particles in the opposing beam that depend nonlinearly on their positions within the bunch. The consequences of this interaction present challenges for the transport and stability of beams within the accelerator, as well as challenges for detector operation due to the generation of secondary particles. The accelerator physics and detector physics challenges are collectively referred to as \emph{beam-beam} (BB) effects~\cite{Herr:1982430, YokoyaChen, item_9f194f9b76e84ba8ab62b19b17597ccc}. Thus, BB modeling is an essential component of any collider design loop.  

BB effects are varied, and their relative importance depends on the collider. 
In electron-positron colliders, as particles traverse the collective field of the opposing bunch, they undergo transverse acceleration and emit synchrotron radiation, referred to as \emph{beamstrahlung}.
This emission increases the energy spread of the beam~\cite{Chen:1987zg}. In a circular machine, the increased energy spread causes the bunch length to increase, and in some cases leads to unstable dynamics~\cite{PhysRevAccelBeams.27.121001}. 
In a linear collider, the beams are dumped after every collision, so beamstrahlung effects are less impactful from a beam dynamics perspective. However, the increased energy spread of the colliding particles broadens the luminosity spectrum, which degrades the performance of colliders optimized for precision physics~\cite{Augustin:1978ah}. 
High energy photons are also generated by the incoherent \emph{radiative Bhabha scattering} process. Bhabha photons carry a large fraction of the radiating particle's energy, and the particle will be lost if its energy is outside the dynamic aperture of the ring. This process is the primary limitation on beam lifetime in high-energy circular colliders~\cite{Burkhardt:1994wk, PhysRevAccelBeams.27.091001}. 
In both linear and circular collider designs, the emitted radiation can generate \emph{secondary electron-positron pairs} that are a source of significant detector backgrounds~\cite{PhysRevAccelBeams.26.111002, Sailer:1443555} or of additional luminosity~\cite{chigusa2025searcheselectroweakstatesfuture}. 
For circular colliders, where beams travel around a closed orbit and are reused every turn, BB interactions are coupled with other effects in the machine including space charge~\cite{PhysRevAccelBeams.27.101001}, impedances~\cite{Zhang_2024, PhysRevAccelBeams.28.051002}, and lattice nonlinearities coming from magnet misalignments or field errors~\cite{zhou:ipac15-weyb3, PhysRevAccelBeams.27.091001}. Together, these effects result in a rich and complex nonlinear dynamics landscape.  

Numerical simulations are essential to model BB effects. 
Because of the strong nonlinearities in the physics at the IP, the predictive power of analytical models is limited to low-order approximations and, in most cases, self-consistent simulations are necessary. The collider community distinguishes between \emph{weak-strong} and \emph{strong-strong} models.
In the former, the \emph{weak} beam evolves under the influence of a \emph{strong} unperturbed beam, while in the latter the two beams evolve self-consistently. In addition, a soft-Gaussian approximation can be used in strong-strong models by assuming that the beam densities are Gaussian and the beam fields are described by the Bassetti-Erskine formula~\cite{BassettiErskine}. 
These approximations may not be sufficient to capture non-equilibrium density profiles that occur at injection~\cite{Xu_EIC_injection} or upon collision~\cite{Babaev2024}.  

Future collider studies demand fast, high-fidelity modeling tools that evolve and are sustained by a dedicated community of software developers~\cite{snowmass21}.
Two of the most established codes for strong-strong BB simulations are GUINEA-PIG~\cite{guineapig} and CAIN~\cite{cain}.
Both have been used in the study and design of colliders for three decades~\cite{Barklow99, SchultePhD} and are trusted by the HEP community. However, they are not designed to run on modern \emph{high-performance computing} (HPC) architectures. 
These codes are based on the \emph{particle-in-cell} method (PIC)~\cite{BirdsallLangdon,HockneyEastwood,DawsonRMP}. 
In PIC simulations, the beam dynamics is modeled using a kinetic plasma description. Specifically, the beam particles are sampled in phase space using a collection of computational particles, or macroparticles, each usually representing many physical particles. The self-consistent EM fields are computed on a grid by solving Maxwell's or Poisson's equations. Quantum electrodynamics (QED) effects are incorporated using Monte Carlo methods. 

In this paper, we introduce the open-source, high-performance, multiphysics PIC code WarpX~\cite{WarpX} as a new tool for beam-beam simulations relevant to future colliders. Compared with legacy codes, WarpX offers several advantages in terms of sustainability, portability, and computational performance~\cite{warpxgordonbell}. These include GPU-accelerated HPC through the AMReX framework~\cite{AMReX_IJHPCA1,AMReX_JOSS}; portability across platforms ranging from laptops to supercomputers, across CPU and GPU vendors, and operating systems; systematic documentation; and an open, active, multidisciplinary community of developers and maintainers.
WarpX is part of the Beam, Plasma \& Accelerator Simulation Toolkit (BLAST)~\cite{HueblAAC22} and can interface with other modeling software, such as particle tracking codes. We adapted WarpX and implemented features essential to BB studies, including beamstrahlung, radiative Bhabha emission, coherent and incoherent pair production, and dedicated luminosity diagnostics. 
In view of the long timescales and unexplored regimes associated with future colliders, such a framework can help ensure that future studies remain reproducible, extensible, and accessible to new generations of users. 

This paper is organized as follows: Section~\ref{sec:numericalmethods} describes the relevant numerical methods; ~\ref{sec:collisionscenarios} introduces the selected collider scenarios; ~\ref{sec:results} presents WarpX convergence studies (version 26.09~\cite{warpx2609}), benchmarks against GUINEA-PIG (C++ version 1.2.2~\cite{guineapig}) and CAIN (version 2.42~\cite{cain}), and computational performance; and~\ref{sec:conclusions} summarizes the main conclusions and future directions. The details necessary to reproduce the results reported in this paper are openly available in a dedicated repository~\cite{warpx4beambeam_zenodo}.

\section{\label{sec:numericalmethods}Numerical methods}
The particle-in-cell (PIC) method is a generic numerical approach for modeling the self-consistent, kinetic, relativistic dynamics of plasmas~\cite{BirdsallLangdon,HockneyEastwood,DawsonRMP}. In this work, we use WarpX as a PIC framework for strong-strong beam-beam simulations that capture the coupled nonlinear evolution of both colliding beams. The two colliding beams are sampled in 6D phase space. A bunch population of $N$ particles is represented by an ensemble of $N_m$ \emph{macroparticles}, each representing $w=N/N_m$ real particles ($w$ is the macroparticle's \emph{weight}). The macroparticles are advanced in time according to the relativistic equations of motion for the position ($\textbf{x}$) and momentum ($\textbf{p}$) under the action of the Lorentz force due to the self-consistent, collective electric ($\textbf{E}$) and magnetic ($\textbf{B}$) fields: 
\begin{equation}
\frac{d \textbf{x}}{d t } = \frac{\textbf{p}}{m \gamma}, \qquad \frac{d \textbf{p}}{dt} = q \left(\textbf{E} + \frac{\textbf{p}}{m \gamma } \times \textbf{B}\right),
\label{eq:partmotion}
\end{equation}
where $\gamma = \sqrt{1 + |\textbf{p}|^2/(m c)^2}$ is the Lorentz factor of the macroparticle, $m$ its mass, and $c$ the speed of light in vacuum. The EM fields are computed on a 3D spatial grid that is fixed in time and common to both beams. In a fully self-consistent electromagnetic treatment, $\textbf{E}$ and $\textbf{B}$ are obtained by solving Maxwell's equations. For ultra-relativistic beams, however, it is advantageous to adopt the electrostatic approximation in the reference frame co-moving with the centroid of each beam. Specifically, in that frame, the electrostatic potential $\phi$ satisfies Poisson's equation, $\Delta \phi = - \rho/ \epsilon_0$, where $\rho$ is the beam charge density and $\epsilon_0$ is the vacuum permittivity. Expressed in laboratory-frame variables, the equation for $\phi$ becomes~\cite{JLV2008}
\begin{equation}
  \left[  \boldsymbol{\nabla}^2 - (\boldsymbol{\beta} \cdot \boldsymbol{\nabla})^2 \right] \phi = - \frac{\rho}{\epsilon_0},
  \label{eq:poisson}
\end{equation} 
where $\boldsymbol{\beta}$ is the average beam velocity in the laboratory frame, normalized to $c$. The electric and magnetic fields are then reconstructed from $\phi$ as 
\begin{equation}
\textbf{E} = - \boldsymbol{\nabla} \phi + \boldsymbol{\beta} (\boldsymbol{\beta} \cdot \boldsymbol{\nabla} \phi), \qquad \textbf{B} =  - \boldsymbol{\beta} \times \boldsymbol{\nabla} \phi / c.
\label{eq:EBfromPhi}
\end{equation}
The particle and field equations -- Eqs.~\ref{eq:partmotion},~\ref{eq:poisson} and~\ref{eq:EBfromPhi} -- are nonlinearly coupled through the Lorentz force acting on the macroparticles and through the charge density deposited on the grid. At each timestep, the charge density is deposited onto the grid, the equation for the electrostatic potential is solved on the grid, the corresponding fields are interpolated to the macroparticles' positions, and the macroparticles are advanced according to the relativistic equations of motion. The PIC algorithm thus evolves particles and fields self-consistently over the duration of the collision.

The PIC algorithm, which describes the deterministic evolution of particles and fields, can be extended to include stochastic processes. Relevant QED processes are incorporated into the PIC loop using Monte Carlo (MC) methods. In the context of collider physics, they can be divided into \emph{coherent} and \emph{incoherent} processes. The former are interactions between the particles of one beam and the collective field of the opposing beam. The latter are interactions between individual particles, real photons, and virtual photons. Coherent processes are modeled within strong-field QED (SF-QED) using the locally constant field approximation (LCFA) (see Refs. \cite{ritus1985quantum,dipiazza.rmp.2012,gonoskov.rmp.2022,fedotov.pr.2023} and references cited therein), while incoherent processes are described using binary-collision MC models~\cite{higginson2020corrected}. The quantum beam-beam processes considered in this work are summarized in Table~\ref{tab:bbprocesses}. The implementation of the coherent processes is described in Ref.~\cite{fedeli2022picsar}. The incoherent processes are modeled using the binary-collision algorithm of Ref.~\cite{higginson2020corrected}, in which one or both incoming particles are, where needed, equivalent virtual photons~\cite{LandauVol4}. 

\begin{table*}
\begin{ruledtabular}
\begin{tabular}{lll}
\textbf{Process} & \textbf{Physical channel} & \textbf{Numerical model} \\
\hline
Beamstrahlung & $e^\pm + n\omega \rightarrow e^\pm + \gamma$ & SF-QED (LCFA) \\
Coherent pair production & $\gamma + n\omega \rightarrow e^+ + e^-$ & SF-QED (LCFA) \\
Compton scattering \, & $e^\pm + \gamma \rightarrow e^\pm + \gamma $ & MC with Klein-Nishina \\
Radiative Bhabha scattering \, & $e^+ + e^- \rightarrow e^+ + e^- + \gamma$ & MC via $e^\pm + \gamma^\ast \rightarrow e^\pm + \gamma$ \\
Linear Breit-Wheeler & $\gamma + \gamma \rightarrow e^+ + e^-$ & MC \\
Bethe-Heitler & $\gamma + e^\pm \rightarrow e^\pm + e^+ + e^-$ & MC via $\gamma + \gamma^\ast \rightarrow e^+ + e^-$ \\
Landau-Lifshitz & $e^\pm + e^\pm \rightarrow e^\pm + e^\pm + e^+ + e^-$ & MC via $\gamma^\ast + \gamma^\ast \rightarrow e^+ + e^-$ \\
\end{tabular}
\end{ruledtabular}
\caption{\label{tab:bbprocesses}
beam-beam processes considered in this work and their numerical treatment.
SF-QED stands for strong-field QED, LCFA for locally constant field approximation~\cite{ritus1985quantum}, MC for Monte Carlo, $\gamma^\ast$ denotes a virtual photon used in the Weizsacker-Williams approximation~\cite{Weizscker1934}, $\gamma$ a real photon, and $n \omega$ many photons from the collective field of the beam.}
\end{table*}

\begin{figure}
    \includegraphics[width=\linewidth]{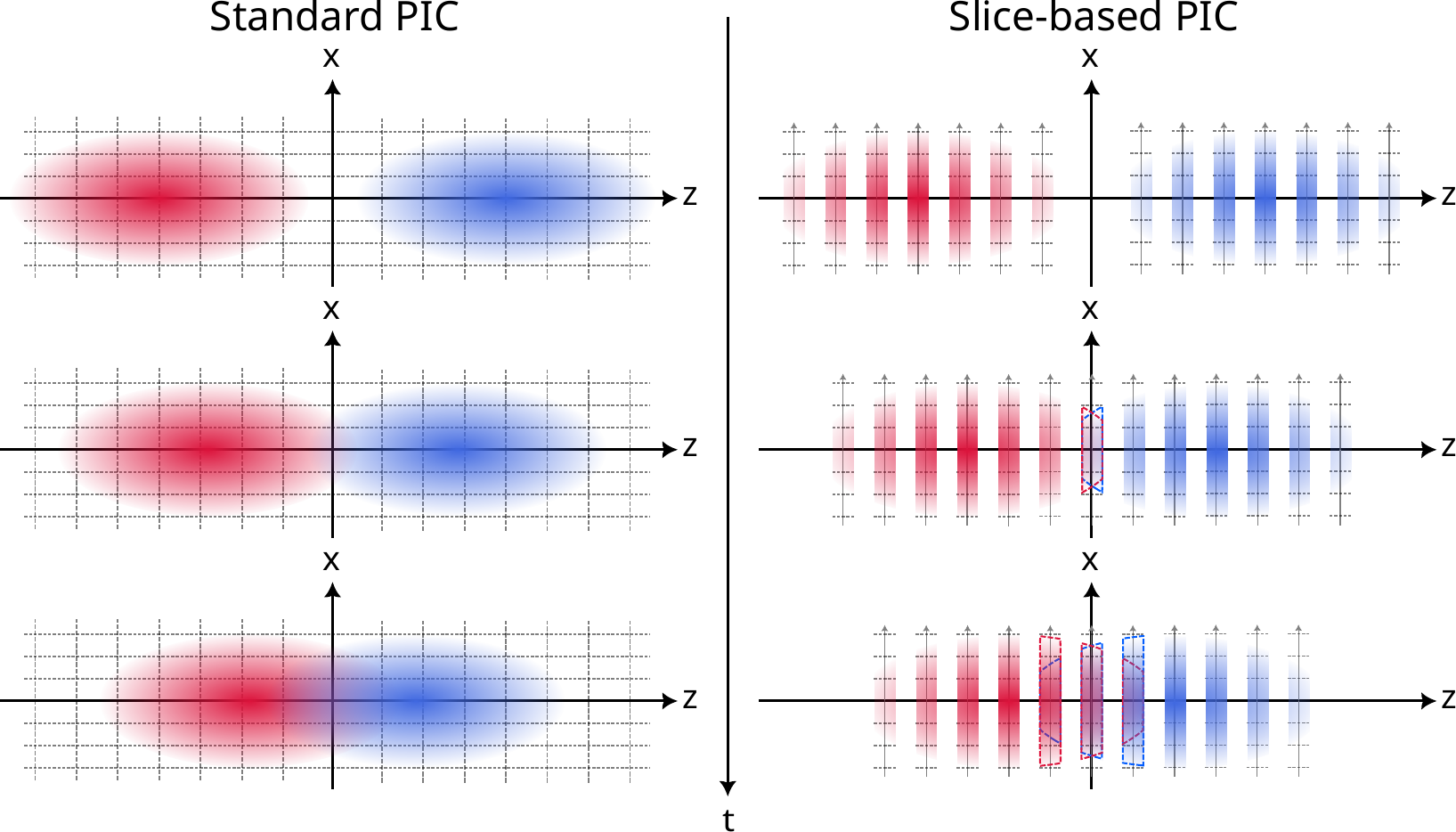}
    \caption{Left: generic PIC evolution of the beam-beam collision on a single, stationary 3D grid. Right: beam-beam slicing approach with sequential slice-to-slice interactions and a co-moving 2D transverse grid per slice.}
    \label{fig:sketch}
\end{figure}

The main algorithmic difference between legacy beam-beam codes like GUINEA-PIG and CAIN with respect to WarpX is in the way the collision advances in time, as sketched in Fig.~\ref{fig:sketch}. GUINEA-PIG and CAIN are specialized and efficient for collider IP studies. They describe the beams as sequences of longitudinal slices and model the collision as a succession of slice-to-slice interactions on transverse grids. The space-charge field is calculated independently on each interacting slice using a 2D Poisson solve. This approximation is justified for ultra-relativistic beams, for which the longitudinal derivatives in Eq.~\ref{eq:poisson} are suppressed by a factor \(1/\gamma^2\). In this formulation, the longitudinal slicing of the beams also fixes the temporal resolution, while processes that are not naturally tied to the slice decomposition (e.g. incoherent pair production) may require shorter time steps. To overcome this potential issue, GUINEA-PIG implements a sub-stepping strategy, although not self-consistently. More details on the numerical algorithms implemented in GUINEA-PIG are described in Appendix~\ref{sec:app_gpalgo}. 

WarpX, by contrast, provides a more general PIC framework in which both beams are evolved self-consistently on a 3D grid that is stationary and common to both beams. Therefore, all PIC operations (charge deposition, field solve, field gathering, particle push) as well as QED evaluations are performed in the whole grid at all time steps, not only on interacting slices. This enables a unified treatment of collective fields, particle dynamics, and stochastic processes within a single simulation loop. For Poisson equation, we implemented two different FFT-based methods that are applied to each beam: a single fully 3D solve~\cite{IGFsolver3d} and many independent 2D solves~\cite{quiang_igf2d}, one for each $z$ grid coordinate. We will use the latter throughout this paper to more closely match the benchmarking codes. 
More details on the numerical algorithms implemented in WarpX that are relevant for this work are described in Appendix~\ref{sec:app_wxalgo}. 


\section{\label{sec:collisionscenarios}Collision scenarios}

Here we illustrate the collision scenarios that we selected for the benchmarks reported in Sec.~\ref{sec:benchmarks}. We consider a set of cases that span the parameter space relevant to established $e^+e^-$ collider designs and exploratory extreme-field beam-beam studies. The main parameters are summarized in Table~\ref{tab:colliders}. 

\begin{table*}
\begin{ruledtabular}
\begin{tabular}{lcccccc} 
\textbf{Beam Parameters} &
\textbf{SuperKEKB} &
\textbf{FCC-ee Z} &
\textbf{nano-QED} &
\textbf{FCC-ee} $t\bar{t}$ &
\textbf{ILC} &
\textbf{LWFA} \\
\hline
energy, $E$ [GeV] &
7.007/4.000 &
45.6 &
125 &
182.5 &
125 &
5000 \\
\hline
intensity, $N$ [$10^{10}$] &
6.531/9.043 &
20.20 &
0.0874 &
13.8 &
2.00 &
0.212 \\
\hline
length, $\sigma_z$ [$\mu$m] &
5000/6000 &
$16.7\times 10^3$ &
$10^{-2}$ &
2370 &
300 &
5 \\
\hline
horiz. size at IP, $\sigma^*_x$ [nm] &
10724/10119 &
7993.75 &
10 &
43577.52 &
516 &
1.75 \\
\hline
vert. size at IP, $\sigma^*_y$ [nm] &
62.16/48.30 &
35.20 &
10 &
25.92 &
7.7 &
1.75 \\
\hline
horiz. norm. emittance, $\epsilon_{n,x}$ [$\mu$m] &
63.1/25.0&
63.4 &
0 &
$754$ &
5 &
0.050 \\
\hline
vert. norm. emittance, $\epsilon_{n,y}$ [nm] &
176.6/67.63 &
157.95 &
0 &
171.39 &
35 &
50 \\
\hline
relative energy spread, $\delta$ [$10^{-3} E$] &
0.637/0.792 &
1.34 &
0 &
1.64 &
1 &
0 \\
\hline
crossing angle, $\theta$ [mrad] &
83 &
30 &
0 &
30 &
0 &
0
\end{tabular}
\end{ruledtabular}
\caption{Summary of the $e^+e^-$ beam parameters used for the various test cases. The SuperKEKB column contains the parameters for the high- and low-energy rings (HER/LER)~\cite{Zhou_2023_PRAB}. FCC is the Future Circular Collider~\cite{benedikt2025future}, ILC the International Linear Collider. The nano-QED case is inspired by Ref.~\cite{Yakimenko}. The LWFA case is representative of a 10 TeV collider based on laser wakefield acceleration~\cite{Schroeder_2023}, of which we also consider a $\gamma\gamma$~\cite{bulanov.arxiv.2026} version.}
\label{tab:colliders}
\end{table*}

SuperKEKB~\cite{Zhou_2023_PRAB} is a circular $e^+ e^-$ collider operating at the luminosity frontier where asymmetric beams collide at GeV-scale energies. SuperKEKB incorporates several design elements of the Future Circular Collider (FCC-ee)~\cite{benedikt2025future}, including a large crossing angle and the crab-waist scheme~\cite{Zobov_2016}. Thus, beam dynamics studies performed with this machine may provide useful insights into the design and optimization of the FCC-ee~\cite{Zimmermann:2025rws}. Two operating points at FCC-ee (Z and $t \bar t$) are considered as well as a case representative of the International Linear Collider (ILC)~\cite{ILCtdr}.
The nanometric QED-dominated (nano-QED)~\cite{Yakimenko} and laser-wakefield accelerator (LWFA)~\cite{gessner2025design} cases probe regimes in which SF-QED physics becomes prominent, owing to extreme bunch compression in the former and multi-TeV beam energies in the latter. The idealized nano-QED case has no momentum spread and exactly spherical beams. 


\begin{figure}
    \includegraphics[width=\linewidth]{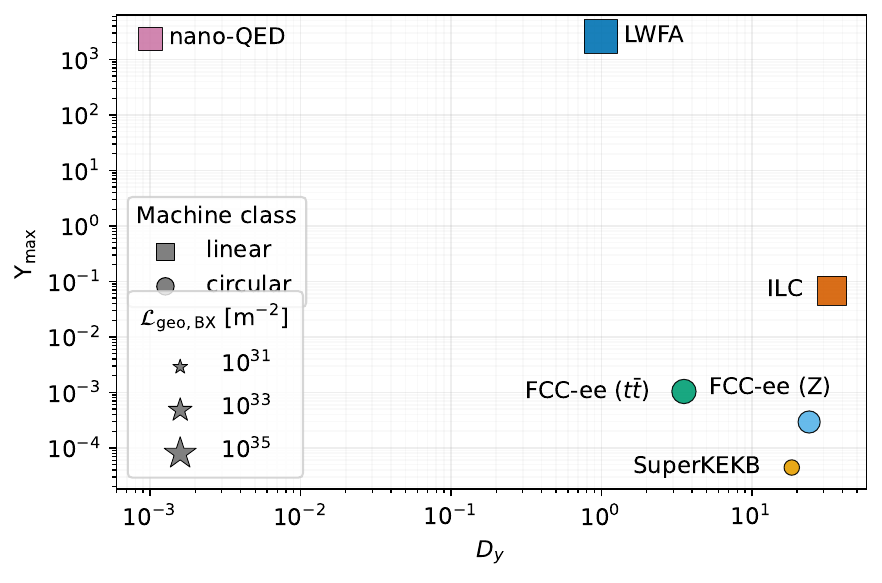}
    \caption{Parameter space explored with the collision cases benchmarked in this paper. The horizontal axis measures the strength of beam-beam effects with the vertical disruption ($D_y$, Eq. \ref{eq:B}), while the vertical axis measures the magnitude of nonlinear QED effects with the beamstrahlung parameter ($\chi$, Eq. \ref{eq:upsmax}). The size of the marker is proportional to the logarithm of the geometric luminosity per bunch crossing ($\mathcal{L}_{\rm geo, BX}$, Eq. \ref{eq:Lgeo}).}
    \label{fig:paramspace}
\end{figure}

These cases span a broad range of beam-beam regimes: spherical vs. round vs. flat beams, crossing angle vs. head-on collisions, long high-charge beams vs. short low-charge beams, linear vs. nonlinear QED physics, and symmetric vs. asymmetric configurations. To visualize the diversity of these regimes in a compact way, Fig.~\ref{fig:paramspace} places the selected scenarios in a three-parameter space defined by a measure of collective beam-beam strength, \(D_y\), a measure of strong-field QED effects, \(\Upsilon_{\max}\), and the nominal geometric luminosity per bunch crossing, \(\mathcal{L}_{\rm geo, BX}\). Specifically, we define the following quantities for each beam, identified by subscripts $i \in \{ 1,2\}$ and, when needed, by coordinate $u\in \{ x,y\} $: \(N_i\) is the bunch population; \(\sigma^*_{u,i}\) are the transverse rms bunch sizes at the IP;  $\beta^*_{u,i}$ are the beta functions at the IP; $\sigma_{z,i}$ is the rms bunch length; $\gamma_i$ is the Lorentz factor associated to the beam energy; \(\theta\) is the half crossing angle; \(r_e\) is the classical electron radius; and $\alpha$ is the fine structure constant. 

$D_{y,\,1\leftarrow 2}$ is the disruption parameter of beam 1 due to beam 2, defined as~\cite{YokoyaChen}:
\begin{equation}
   D_{y,\,1\leftarrow 2} = \dfrac{2 N_2 r_e \ell_{int}}
{\gamma_1 \sigma_{y2}\left(\sigma_{x2}+\sigma_{y2}\right)}, 
\label{eq:B}
\end{equation}
where $\ell_{int}$ is a measure of the interaction or overlap region. 
For head-on collisions, $\ell_{\rm int}=\sigma_{z2}$. For collisions with a horizontal crossing angle without crab-waist optics, we use an effective overlap length $\ell_{\rm int}\simeq \sigma_{z2}/\sqrt{1+\phi_2^2}$, where $\phi_2=\theta\sigma_{z2}/\sigma_{x2}^*$ for half crossing angle $\theta$. For crab-waist schemes (i.e. SuperKEKB and FCC-ee) the relevant interaction length is instead set by the shortened overlap region, of order $\sigma_x^*/\theta$ and comparable to the design $\beta_y^*$.

$\Upsilon_{\max}$ is the maximum beamstrahlung parameter experienced by beam 1 in the field of beam 2, defined as~\cite{YokoyaChen}:
\begin{equation}
\Upsilon_{\max, 1\leftarrow 2} = \frac{2\,r_e^2 \gamma_1 N_2}
{\alpha\,\sigma_{z2}\left(\sigma_{x2}+\sigma_{y2}\right)}
\label{eq:upsmax}    
\end{equation}

The geometric luminosity per bunch crossing is computed from the space-time overlap of the two bunch densities,
\begin{equation}
\mathcal{L}_{\rm geo, BX}
=
K \int dt \int d^3{\bf r}\,
n_1({\bf r},t)n_2({\bf r},t),
\label{eq:Lgeo}
\end{equation}
where \(K\) is the M{\o}ller velocity factor~\cite{furman2003moller}, with \(K\simeq 2c\) for ultrarelativistic head-on beams. For Gaussian bunches, including unequal beam sizes, a horizontal crossing angle, and the hourglass effect, this can be written in the compact form as follows
\begin{equation}
\mathcal{L}_{\rm geo, BX}
=
\frac{N_1N_2}{2\pi}
\!\!\int_{-\infty}^{+\infty} ds \, \frac{\Lambda(s)\exp \!\left[
-2 \left( \frac{s\tan \theta}{\Sigma_x(s)}\right)^2
\right]}
{\Sigma_x(s)\Sigma_y(s)} ,
\label{eq:Lgeo_compact}
\end{equation}
with
\begin{equation}
\Sigma_{x}^2(s)=\sigma_{x1}^2(s)+\sigma_{x2}^2(s),
\quad 
\Sigma_{y}^2(s)=\sigma_{y1}^2(s)+\sigma_{y2}^2(s),
\end{equation}
and
\begin{equation}
\sigma_{ui}(s)
=
\sigma_{ui}^*
\sqrt{
1+\left(\frac{s}{\beta_{ui}^*}\right)^2
\label{eq:sigmas}
}, 
\qquad u=x,y; \; i=1,2.
\end{equation}
Here \(\Lambda(s)\) is the normalized longitudinal overlap distribution, \(\int ds\,\Lambda(s)=1\). For Gaussian bunch lengths,
\begin{equation}
\Lambda(s)
=
\frac{1}{\sqrt{2\pi}\sigma_L}
\exp\left[-\frac{s^2}{2\sigma_L^2}\right],
\quad
\sigma_L^2
=
\frac{\sigma_{z1}^2\sigma_{z2}^2}
{\sigma_{z1}^2+\sigma_{z2}^2}.
\end{equation}
The $s$-dependence of \(\sigma_{ui}(s)\), with the beam waists taken at the IP ($s=0$), accounts for the hourglass effect, while the term \(2s\tan\theta\simeq 2 s \theta\) gives the geometric reduction from the finite crossing angle. In the symmetric, head-on, no-hourglass limit this expression reduces to \(\mathcal{L}_{\rm geo, BX}=N_1N_2/(2\pi\Sigma_x\Sigma_y)\).
Figure~\ref{fig:paramspace} shows how the collision scenarios considered in this paper cover the relevant parameter space.

\begin{figure*}
    \includegraphics[width=\linewidth]{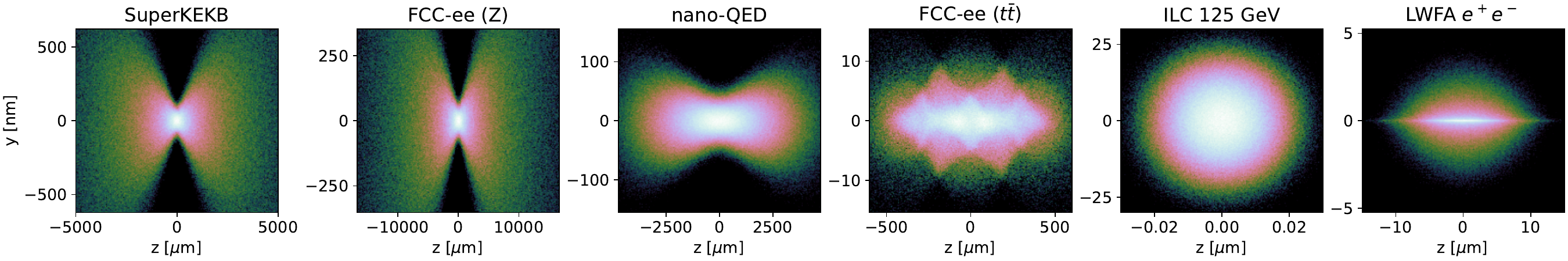}
    \caption{Visual comparison of the product of the two beam densities projected onto the $(z,y)$ plane at the time of maximum overlap for the selected cases, as a qualitative proxy of the spatial distribution of luminosity. Color scale and axis limits differ between panels.}
    \label{fig:lumiproxy}
\end{figure*}

A qualitative view of the collision is given in Figure~\ref{fig:lumiproxy}. It shows, for the six scenarios, a map of the product of the two beam densities projected onto the $(z,y)$ plane at the time of maximum overlap obtained with WarpX. These snapshots highlight the large differences in collision geometry, morphology, aspect ratio, and spatial scale across the selected cases. These cases test both physical fidelity and numerical robustness across very different conditions in terms of beam-beam and QED effects, and nominal luminosity. 

\section{\label{sec:results}Results}
In this section, we discuss the results of WarpX (version 26.09~\cite{warpx2609}) convergence studies, cross-code comparisons with GUINEA-PIG (C++ version 1.2.2~\cite{guineapig}) and CAIN (version 2.42~\cite{cain}), and scaling performance. The baseline numerical parameters used for the benchmarks are summarized in Table~\ref{tab:base_params}. The complete input files are archived in a dedicated repository~\cite{warpx4beambeam_zenodo}.

The baseline parameters are chosen to match the discretizations of the two codes as closely as possible, given the different numerical descriptions of the collision. WarpX evolves both beams on a single fixed 3D grid, whereas GUINEA-PIG computes successive slice-to-slice interactions. Thus, we use $N_z$ cells along the longitudinal direction in WarpX and $N_z/2+1$ slices for each beam in GUINEA-PIG; the additional slice is used to exactly match the number of WarpX time steps. 

The simulation domain, $L_x \times L_y \times L_z$, is chosen to contain the truncated bunch distributions at initialization, while avoiding initial overlap. We consider Gaussian beams truncated at $4 \sigma$ in each direction. With this convention, the centroids are initialized at $z = \pm s_0$, where $s_0 = L_z /4$. The transverse beam sizes at initialization, $\sigma_{ui}(s_0)$ with $u=x,y$, are determined by the hourglass effect, according to Eq.~\ref{eq:sigmas}.

For the SuperKEKB and FCC-ee cases, which include a finite crossing half-angle $\theta$, the simulations are performed in an approximate boosted frame~\cite{Hirata_1995_APS} in which the particle positions are rotated while the beams collide head-on. The details of this approximation are explained in Appendix~\ref{sec:app_boost}. This treatment is appropriate for small crossing angles and ultrarelativistic beams, for which the longitudinal field components are strongly suppressed. The crab-waist transformation is not included in the present benchmarks. The remaining scenarios are modeled as simple head-on collisions. For ILC, this is consistent with crab-cavity compensation~\cite{VERDUANDRES2016193}. 

A general expression for the domain size is:
\begin{align}
    L_x & = 2 n_{\mathrm{cut},x} \max_{i=1,2} ( \sigma_{xi}(s_0)\cos\theta + \sigma_{zi} \sin\theta)  \nonumber \\
    L_y  & = 2 n_{\mathrm{cut},y} \max_{i=1,2} \sigma_{yi}(s_0), \label{eq:box}\\
    L_z & = 16 \max_{i=1,2}  \sigma_{zi} \cos\theta.  \nonumber 
\end{align}
The parameter $n_{\mathrm{cut}, u}$ controls the margin between the truncated initial bunch distribution and the computational boundary. 
The minimal domain, obtained with $n_{\mathrm{cut}, u}=4$, reduces to $L_x = 8\sigma_{x}(s_0)$, $L_y = 8\sigma_{y}(s_0)$, and $L_z = 16\sigma_{z} $ for symmetric beams and zero crossing angle. Cases with strong beam disruption or significant secondary-particle production require larger transverse domains, because disrupted beam particles and secondaries can move beyond the initial $4\sigma$ envelope. We use different $n_{\mathrm{cut}, u}$ depending on the collision scenario~\cite{warpx4beambeam_zenodo}.
In all cases, the beams are evolved for $T=L_z/(2c)$.
The resolution of the baseline simulations is selected upon performing the convergence study reported below.  

\begin{table}
\begin{ruledtabular}
\begin{tabular}{c|c} 
\multicolumn{2}{c}{\textbf{Baseline simulation parameters}}\\ [0.5ex] 
\hline
$L_x \times L_y \times L_z$ & from Eq.~\ref{eq:box}, with case-dependent $n_{\mathrm{cut}, u} \geq 4$\\
\hline
$T$ &   $L_z / (2 c) $  \\
\hline
$N_x \times N_y \times N_z$ &  $512 \times 512 \times 512 $\\
\hline
$N_t$ & $N_z$  \\
\hline
$N_m$ & $ 10^6$ per beam \\
\hline 
Exceptions & $N_y = 2048$ for SuperKEKB and FCC-ee Z \\ 
 & $N_y = 1024$ for ILC \\ 
 & $N_m = 5 \times 10^5$ for ILC and LWFA \\
\end{tabular}
\end{ruledtabular}
\caption{Summary of the baseline simulation parameters used in the benchmarks. Case-specific details are reported in the input files~\cite{warpx4beambeam_zenodo}.}
\label{tab:base_params}
\end{table}

\subsection{\label{sec:convergence}Convergence}

\begin{figure*}
    \includegraphics[width=\linewidth]{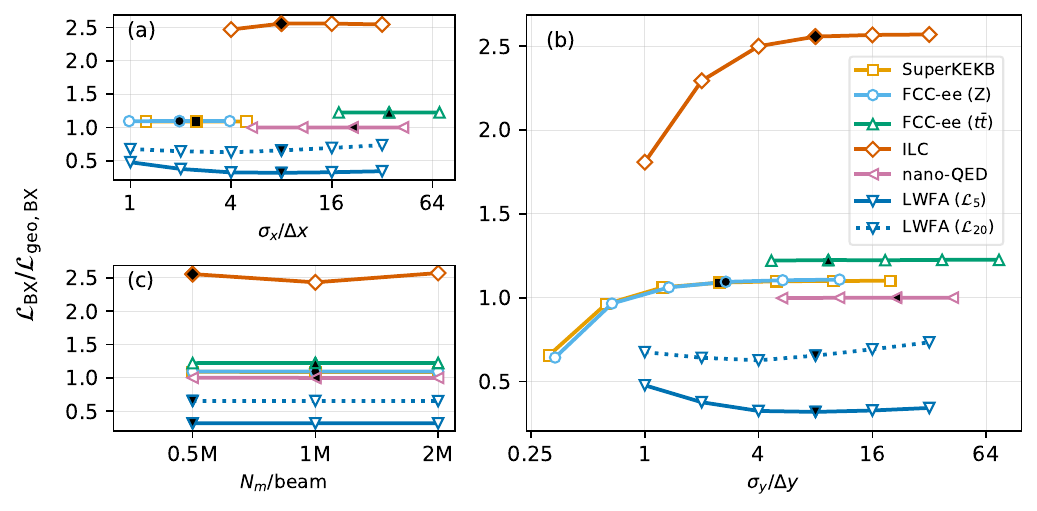}
    \caption{Convergence scans for each collider case. The luminosity per bunch crossing, $\mathcal{L}_{\rm BX}$, is normalized to the geometric value, $\mathcal{L}_{\rm geo,BX}$, and shown as a function of (a) normalized horizontal resolution $\sigma_x^*/\Delta x$, (b) normalized vertical resolution $\sigma_y^*/\Delta y$, and (c) macroparticles per beam $N_m$. For round beams, $\sigma_y^*/\Delta y = \sigma_x^*/\Delta x$. Black-filled markers indicate the baseline simulations used in the cross-code benchmarks. For the LWFA case, $\mathcal{L}_5$ (and $\mathcal{L}_{20}$, dotted line), the luminosity integrated above 95\% (and 80\%) of the nominal center-of-mass energy, is used instead of the total luminosity.}
    \label{fig:conv}
\end{figure*}

Figure~\ref{fig:conv} summarizes the resolution scans for the six collision scenarios introduced in Table~\ref{tab:colliders}.
It shows the numerical convergence of the normalized luminosity per bunch crossing $\mathcal{L}_{\rm BX}/\mathcal{L}_{\rm geo,BX}$ as a function of (a) horizontal grid resolution, $\sigma^*_x/\Delta x$; (b) vertical grid resolution, $\sigma^*_y/\Delta y$; and (c) macroparticle count, $N_m$. We vary one parameter at a time, keeping the others at their baseline values, except for round beams, where we vary the horizontal and vertical grid resolutions together while maintaining $\Delta x=\Delta y$. The longitudinal and temporal resolutions are fixed to $\sigma_z / \Delta z = \sigma_z /(2 c \Delta t) = 32$ for the longer bunch in each case. This should be enough to resolve the longitudinal dynamics, although a dedicated study would be needed depending on the observable of interest. 

Because the physical domain size depends on the collision geometry and beam parameters, the same normalized resolution corresponds to different absolute numbers of grid cells. We vary $N_x$ and $N_y$ (flat) or $N_x = N_y$ (round) from a minimum of $2^6=64$ up to a maximum of $2^{14} = 16384$, depending on the case. The macroparticle count is varied in $N_m \in \{5\times10^5, 10^6, 2\times10^6 \} $. The baseline simulation for each case, which will be used for the benchmarks in Sec.~\ref{sec:benchmarks}, is shown with a black-filled marker.

The scans show a stable behavior of the luminosity when varying the horizontal resolution and the macroparticle count, with the exception of a limited variation observed in the ILC case. Moreover, convergence of the luminosity to within a few percent of the highest-resolution value generally requires resolving one rms vertical IP beam size with several cells, typically $\sigma_u^*/\Delta u \sim 2$--$8$, where $u=y$ for flat beams and $u=x,y$ for round beams. 

The nano-QED case is less demanding because the beams are spherical, therefore the resolution requirements are comparable in all three directions and are reached with relatively few grid cells. SuperKEKB and FCC-ee~Z are the most demanding cases with finite crossing angle. Their long bunches and small $\beta_y^*$ values lead to strong hourglass growth between the IP and the initialization point, so the vertical beam size at initialization is much larger than at the IP. At the same time, the crossing angle increases the required transverse domain through the $\sigma_z\sin\theta$ contribution in $L_x$. In FCC-ee~$t\bar t$, the shorter bunch length and larger $\beta^*$ values reduce the hourglass enlargement between IP and initialization, so the required normalized vertical resolution is reached with fewer cells. The ILC case is also demanding: it is modeled as an effective head-on collision, consistent with crab-cavity compensation, and its very flat, strongly disrupted beams make the luminosity sensitive to all three numerical parameters. For the present scans, numerical convergence requires $N_y\geq 2048$ for SuperKEKB and FCC-ee~Z, and $N_y\geq 1024$ for ILC.

The LWFA case shows yet more complex behavior. In this case, we use the luminosity integrated above 95\% of the nominal center-of-mass energy, $\mathcal{L}_5$, as the primary metric of numerical convergence to focus on the high-energy part of the luminosity spectrum~\cite{Barklow_2023}. We also show $\mathcal{L}_{20}$ (dotted line) to illustrate the sensitivity to the chosen energy threshold. Neither curve is strictly monotonic, but $\mathcal{L}_5$ has a smaller residual variation at high transverse resolution than $\mathcal{L}_{20}$. Between the baseline and finest transverse grids, $\mathcal{L}_5$ and $\mathcal{L}_{20}$ change by approximately 7\% and 11\%, respectively, relative to the finest-grid values. The larger variation in $\mathcal{L}_{20}$ is consistent with its greater sensitivity to lower-energy contributions. At the benchmark resolution, the total luminosity is of order $\mathcal{L}_{\rm BX}/\mathcal{L}_{\rm geo,BX} \sim 100$ (see Fig.~\ref{fig:lumibars}), while $\mathcal{L}_{20}/\mathcal{L}_{\rm geo,BX}\simeq 0.65$. This gap, together with Fig.~\ref{fig:validation_summary}(f), shows that low-energy secondary contributions dominate the total luminosity.

Overall, the baseline grid resolves the luminosity metric to within a few percent of the finest-grid result for the five non-LWFA cases. The LWFA high-energy luminosity remains more sensitive to transverse resolution over the range scanned.

\subsection{\label{sec:benchmarks}Benchmarks}

In this section, we benchmark WarpX against GUINEA-PIG for the collision scenarios introduced in Sec.~\ref{sec:collisionscenarios} and against CAIN in a $\gamma\gamma$ variant of the LWFA collider (see Ref.~\cite{bulanov.arxiv.2026}). We compare luminosity, coherent and incoherent photon emission, and secondary-particle production. These observables test both the self-consistent beam dynamics and the stochastic QED modules. 

\begin{figure}
    \includegraphics[width=\linewidth]{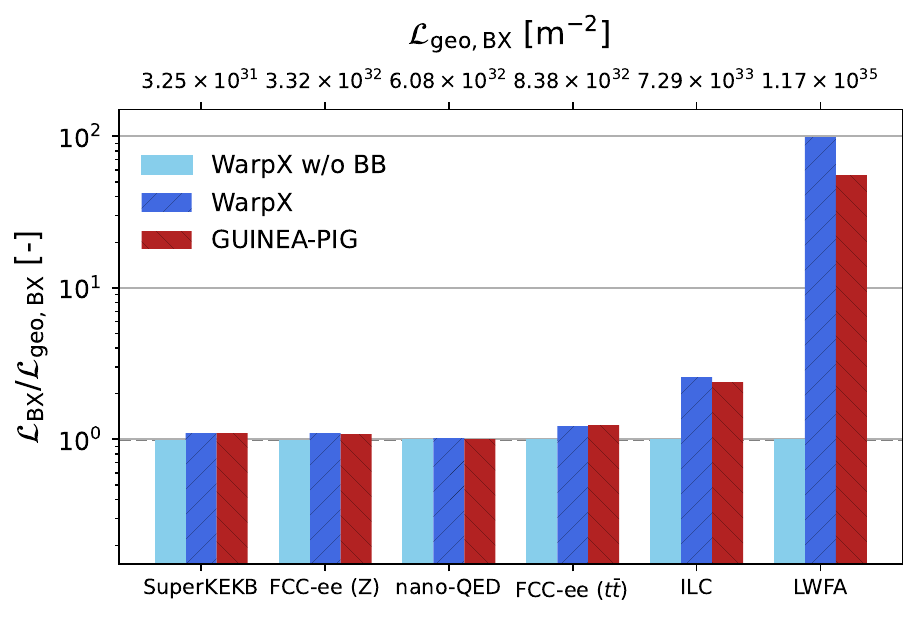}
    \caption{Luminosity per bunch crossing normalized to the geometric value obtained with WarpX (dark blue, striped) and GUINEA-PIG (red, striped). The geometric luminosity is calculated using Eq.~\ref{eq:Lgeo_compact} and reported on the top $x$-axis. WarpX is also run with all BB effects disabled (light blue) to check consistency. The gray dashed line marks $\mathcal{L}_{\rm BX} = \mathcal{L}_{\rm geo, BX}$.}
    \label{fig:lumibars}
\end{figure}

Figure~\ref{fig:lumibars} compares the total luminosity per bunch crossing obtained with WarpX (dark blue, striped) and GUINEA-PIG (red, striped), normalized to the corresponding geometric value. The geometric luminosity is calculated from Eq.~\ref{eq:Lgeo_compact}, using the nominal beam parameters listed in Table~\ref{tab:colliders}. This calculation includes both the crossing angle and the hourglass effect. In addition, WarpX simulations are performed with all BB effects disabled (no self-consistent fields, no beamstrahlung, no secondary-particle generation) to confirm that the luminosity diagnostic recovers the expected geometric overlap (light blue).

The total luminosity can be larger than the geometric luminosity because of the mutual focusing of the two bunches during the collision, i.e. the ``pinch effect''~\cite{YokoyaChen}, or, when secondary encounters are included, because coherent pair production generates copious electron-positron pairs which can themselves contribute to the luminosity. The enhancement is more pronounced for cases with relatively large disruption parameter (ILC) and for the QED-dominated case (LWFA), consistent with the parameter space shown in Fig.~\ref{fig:paramspace}. Slight enhancements are also obtained with FCC-ee and SuperKEKB beams, as also seen in the convergence study in Fig.~\ref{fig:conv}.

WarpX and GUINEA-PIG agree well for cases outside the strongest disruption and QED-dominated regimes (SuperKEKB, FCC-ee Z, FCC-ee~$t\bar t$, and nano-QED), where the relative difference between the two codes remains below 1.3\%. 
For ILC, the total and high-energy luminosities ($\mathcal{L}_5$ and $\mathcal{L}_{20}$) differ by about 8\%.
For LWFA, the total luminosity differs by about 78\% and removing the dominant low-energy contribution still leaves a roughly 40\% difference in $\mathcal{L}_5$. Thus, broadly similar luminosity spectra do not necessarily imply precision agreement in the observables relevant to these colliders. Resolving the differences will require dedicated studies of beam evolution, radiation, pair feedback, and numerical convergence. 

\begin{figure*}
    \includegraphics[width=\linewidth]{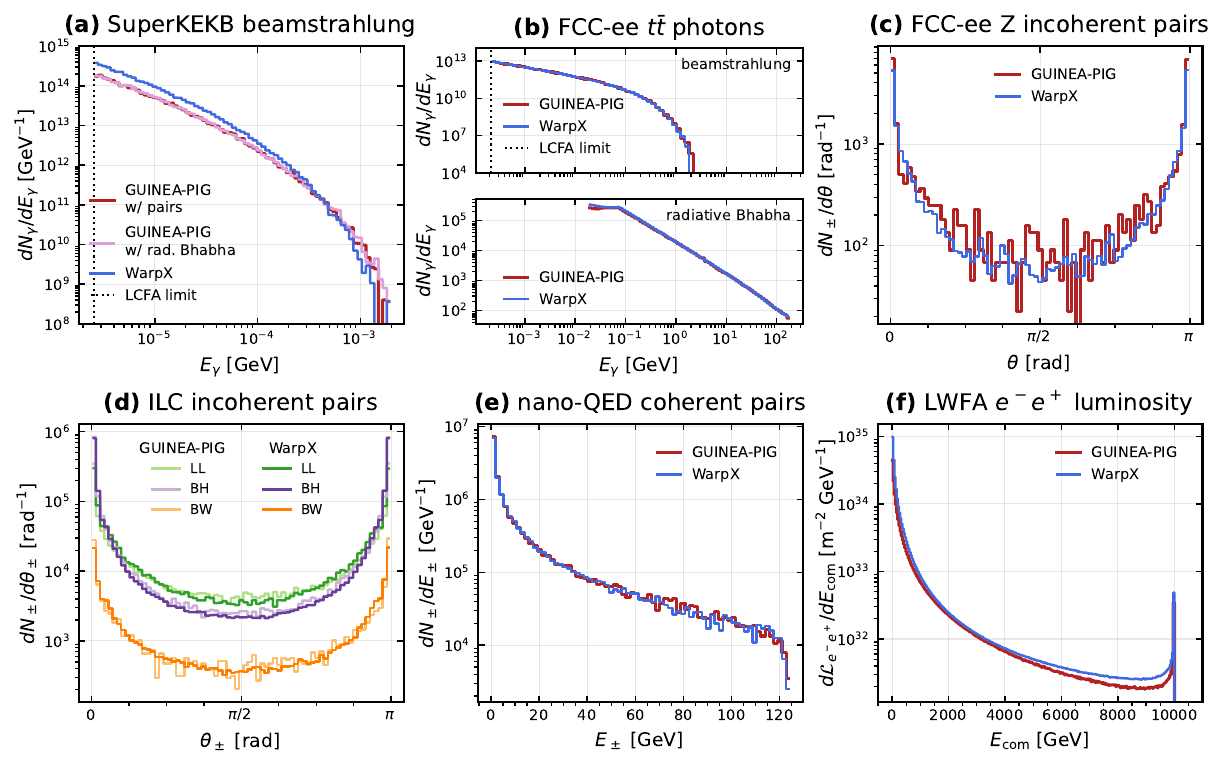}
    \caption{Summary of the WarpX benchmarks against GUINEA-PIG. WarpX results are shown in blue, while GUINEA-PIG results are shown in red, unless otherwise stated.
    (a) SuperKEKB: beamstrahlung photon spectra. GUINEA-PIG is run twice, with incoherent pair creation (red) and radiative Bhabha (pink). WarpX includes both processes. The black dashed line marks the minimum photon energy for LCFA validity~\cite{DiPiazza_2019_PhysRevA}.
    (b) FCC-ee~$t\bar t$: beamstrahlung (top) and radiative Bhabha (bottom) photon spectra.
    (c) FCC-ee~Z: distribution of the polar angle of incoherent electron-positron pairs. 
    (d) ILC: distribution of the polar angle of incoherent electron-positron pairs, separated according to the production channel. Landau-Lifshitz in green, Bethe-Heitler in purple, Breit-Wheeler in orange. WarpX results are shown in dark colors, GUINEA-PIG in light colors.  
    (e) nano-QED: spectra of coherent electron-positron pairs.
    (f) LWFA: differential luminosity due to electron-positron encounters.}
    \label{fig:validation_summary}
\end{figure*}

The agreement between WarpX and GUINEA-PIG is summarized in Fig.~\ref{fig:validation_summary}, which shows selected representative diagnostics for each benchmark scenario. These panels were chosen to sample the main physics modules: beamstrahlung, radiative Bhabha emission, incoherent pair generation, coherent pair generation, and luminosity spectra. The complete set of luminosity, photon, and pair distributions is provided in the archived data.

During the collision, photons can be radiated via coherent or incoherent processes.
Beamstrahlung is relevant in all the simulated cases, while radiative Bhabha is turned off in the nano-QED and LWFA cases, where coherent processes dominate. 
Figure~\ref{fig:validation_summary}(a) shows the beamstrahlung spectra for SuperKEKB obtained with GUINEA-PIG (red, pink) and WarpX (blue). The two GUINEA-PIG runs separately include radiative Bhabha photon emission (orange) and incoherent pair generation (red), while WarpX includes them in the same run. The vertical black dashed line marks the estimated minimum photon energy for LCFA validity, computed from Eq.~(12) of Ref.~\cite{DiPiazza_2019_PhysRevA}. Because the incoherent processes should not feedback into the beamstrahlung spectrum, comparing the two GUINEA-PIG results is a sanity check. Figure~\ref{fig:validation_summary}(b) shows the spectra of the photons emitted via beamstrahlung (top) and radiative Bhabha scattering (bottom) for FCC-ee~$t\bar t$. The spectra agree throughout the full energy range. 

During the collision, secondary electron-positron pairs can be generated via incoherent or coherent processes. 
Incoherent pair creation is turned on for SuperKEKB, FCC-ee, and ILC, where coherent processes are negligible. Instead, coherent pair creation is turned on for the nano-QED and LWFA cases, where it dominates. Figure~\ref{fig:validation_summary}(c) and (d) show the polar-angle distribution of incoherent pairs for FCC-ee~Z and ILC, respectively, while panel (e) shows the coherent pair spectra for the nano-QED case. 

For FCC-ee~Z, the GUINEA-PIG incoherent-pair angular distributions show larger statistical fluctuations than the WarpX distributions. In GUINEA-PIG, the pair statistics are tied to the sampling of the primary beams, i.e. to the number of macroparticles. In WarpX, the reduced beam sampling can be compensated by using a multiplier that increases the collision probability and reduces the weight of the produced particles accordingly~\cite{higginson2020corrected}.
For ILC, the pairs are further separated according to production channel. The agreement is good for the Landau-Lifshitz, Bethe-Heitler, and Breit-Wheeler contributions.

Beamstrahlung and secondary-particle production broaden the $e^-e^+$ luminosity spectra in the ILC and LWFA cases and reduce the peak at the design energy. Figure~\ref{fig:validation_summary}(f) shows the luminosity spectra due to electron-positron encounters for the LWFA case obtained with GUINEA-PIG (red) and WarpX (blue). The dominant contribution is at the lowest energies. In WarpX, approximately 95\% of the integrated $e^-e^+$ luminosity involves at least one secondary electron or positron produced by coherent pair creation, and approximately 68\% comes from encounters between two such secondaries. The overall spectral shapes are similar, but discrepancies occur across the energy range, including near the lowest energies and the design energy. The LWFA case enters a QED-cascade regime~\cite{gonoskov.rmp.2022}, namely a regime with repeated photon emissions, photon conversion to pairs, and emissions by those pairs. Differences in the codes' QED sampling could therefore contribute to the observed discrepancy, although this comparison does not isolate their effect

As a final benchmark, we consider a $\gamma\gamma$ collider based on wakefield technology. 

In this configuration, two accelerated electron bunches undergo a conversion stage followed by an interaction-point collision~\cite{Schroeder_2023,bulanov.arxiv.2026}. 
First, the two electron bunches each collide with a laser pulse at separate conversion points (CPs). At the CPs, the laser photons are Compton-backscattered to high energies and electron-positron pairs can be generated through Breit-Wheeler interactions between laser photons and backscattered photons. The resulting mixture of high-energy photons, electrons, and positrons from the two CPs then collides at the IP.  Here, we benchmark WarpX against CAIN, which was designed to model $\gamma\gamma$ colliders in addition to $e^+e^-$ colliders.

We consider electron beams with the same parameters as in the LWFA collider, i.e. 5 TeV, $2.12 \times 10^9$ particles, $\sigma_z = 5$ $\mu \rm m$, $\sigma_x^* = \sigma_y^* = 1.75 $ nm. The distance from each CP to the IP is chosen as the minimum distance that accommodates the laser pulse: approximately $1.5$ laser lengths plus a few bunch lengths (here equal to 2.43 mm). We use a laser pulse with normalized vector potential $a_0=0.34$, rms length $1.6 \,\mathrm{mm}$, photon energy $2.5\,\mathrm{eV}$, and rms transverse size $20\lambda$, where $\lambda$ is the laser wavelength. The grid spacing and time step are chosen to resolve the electron beam size and bunch duration at the CP and IP. The complete input parameters are provided in the archived data.

The CP is modeled differently in WarpX and CAIN. In WarpX, the laser is represented by a collection of photons and the Compton scattering process is modeled as a binary collision between an electron macroparticle and a laser-photon macroparticle. By contrast, CAIN samples the Compton events using the analytical formula for the laser field. Because the laser is much wider than the electron beam, WarpX samples only the central region overlapping the electron beam. This is done with a photon-flux approximation in which laser focusing is neglected and the laser photons propagate parallel to $z$.

\begin{figure*}
    \includegraphics[width=\linewidth]{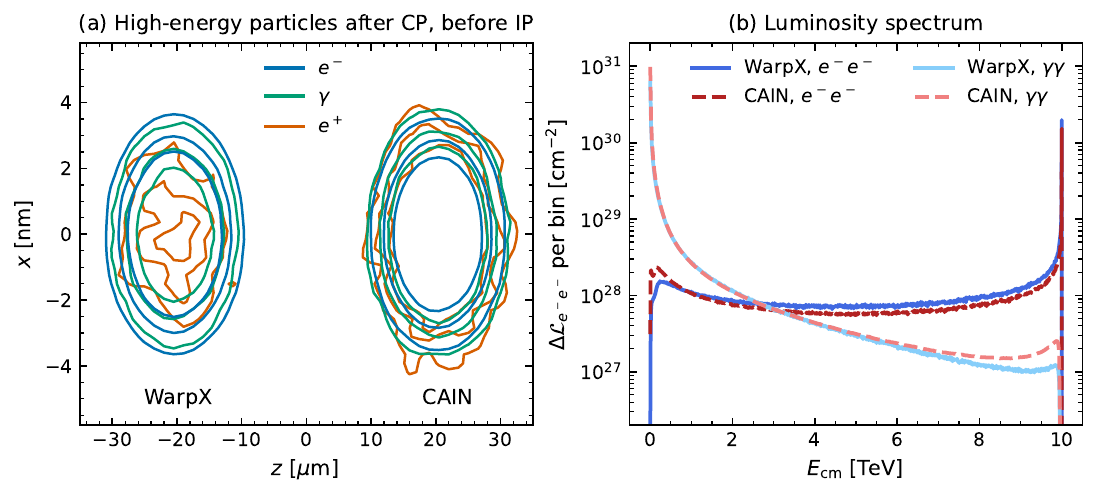}
    \caption{Comparison of WarpX and CAIN for the $\gamma\gamma$ collider configuration.
    (a) Transverse-longitudinal density contours at the IP for high-energy particles with $E>0.95E_0$, where $E_0=5\,\mathrm{TeV}$. Colors indicate particle species: blue for electrons, green for photons, and orange for positrons. WarpX results are shown on the left side, CAIN results on the right. The same contour levels are used for both codes for each species.
    (b) Corresponding $e^-e^-$ and $\gamma\gamma$ luminosity spectra for CAIN (dashed dark and light red lines, respectively) and WarpX (solid dark and light blue, respectively).}
    \label{fig:gammagamma}
\end{figure*}

Figure~\ref{fig:gammagamma}(a) shows the distribution of particles with energy within 5\% of the beam energy at the exit of the CPs and before colliding at the IP obtained with WarpX (left) and CAIN (right). The two codes give similar transverse-longitudinal distributions for the high-energy particles transported from the CPs to the IP. Figure~\ref{fig:gammagamma}(b) shows the luminosity spectra for $e^-e^-$ and $\gamma \gamma$ encounters. The spectral shapes are similar, although the agreement is not exact. CAIN converts more high-energy electrons into high-energy photons. Nonetheless, the level of agreement is satisfactory given that the IP collision depends on the particle distributions produced at the two CPs, which are modeled differently in the two codes. 

\subsection{\label{sec:performance}Performance}
We evaluate the performance and scalability of WarpX on Perlmutter, a supercomputer hosted at NERSC.
The simulations are run on Perlmutter large-memory GPU nodes, each with four NVIDIA A100 GPUs with 80 GB of HBM, one 64-core AMD EPYC 7763 CPU, and 256 GB of DDR4 memory~\cite{nersc_perlmutter}. 

WarpX is designed for exascale HPC architectures~\cite{warpxgordonbell}, enabling simulations with large grids, high macroparticle counts, and substantial generated-particle populations. This capability is particularly relevant for high-disruption or finite-crossing-angle cases, where fine spatial resolution is required (e.g. FCC-ee, SuperKEKB, ILC), and for cascade regimes (e.g. LWFA), where secondary-particle production can rapidly increase the particle count and memory footprint.

Table~\ref{tab:benchmark_timings} shows the timings of the benchmark simulations considered in the previous sections.
The WarpX production runs (performed with version 26.09~\cite{warpx2609}) complete in a few minutes on 4 GPUs, i.e. one Perlmutter GPU node, whereas the corresponding GUINEA-PIG (C++ version 1.2.2~\cite{guineapig}) and CAIN (version 2.42~\cite{cain}) runs require several hours to more than one day on one CPU core.
This is not intended as a hardware-normalized comparison, since GUINEA-PIG and CAIN are serial CPU codes whereas WarpX targets GPU-parallel execution. 
Rather, the comparison reflects the practical turnaround time available to users for production simulations. 
\begin{table}
\begin{ruledtabular}
\begin{tabular}{c|c|c}
Case                  & WarpX time [min] & GUINEA-PIG time [min] \\
\hline
SuperKEKB             & 4.87  & 2169 / 2136 \\
FCC-ee Z              & 5.28 & 2305 / 2353 \\
nano-QED              & 1.90 & 267         \\
FCC-ee $t\bar t$      & 1.70 & 279 /  294  \\
ILC                   & 3.72  & 822 /  837  \\
LWFA $e^+e^-$         & 5.59 & 610         \\
\hline
Case                  & WarpX time [min] & CAIN time [min] \\
\hline
LWFA $\gamma \gamma$  & 6.81  & 2714
\end{tabular}
\end{ruledtabular}
\caption{Wall-clock times for the benchmark simulations. WarpX runs on 4 GPUs, GUINEA-PIG and CAIN on 1 CPU. All timings include beam-beam fields, QED processes, generated secondary particles, and diagnostics, and, for the $\gamma\gamma$ case, the conversion point. When two GUINEA-PIG timings are reported, they correspond to the simulations with either incoherent pair creation or radiative Bhabha scattering enabled, respectively; otherwise, only coherent pair production is enabled. In WarpX, incoherent pair creation and radiative Bhabha scattering are enabled at the same time. The comparison is intended to show practical production turnaround time rather than a hardware-normalized performance benchmark.}
\label{tab:benchmark_timings}
\end{table}

\begin{figure*}
    \includegraphics[width=\linewidth]{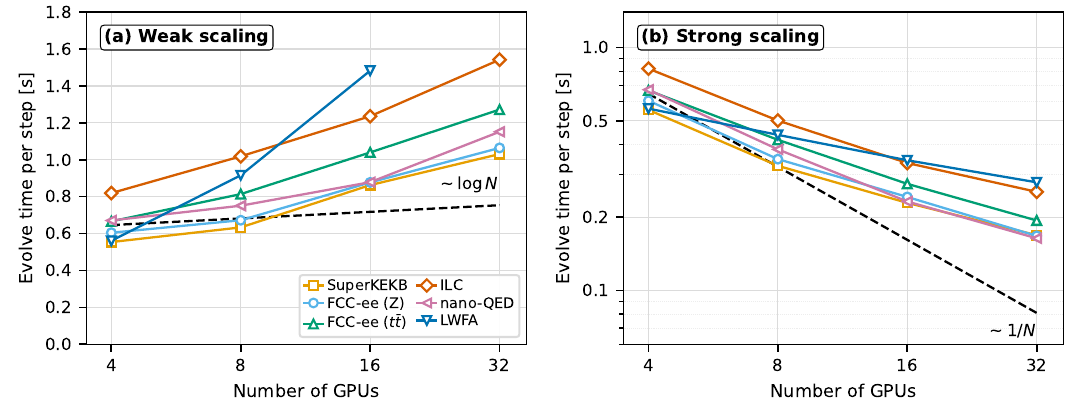}
    \caption{Performance scaling of electrostatic WarpX for production-level beam-beam simulations across the different collision scenarios: (a) weak scaling, where the grid cells and macroparticle count increase proportionally with the number of GPUs to maintain approximately constant workload per GPU, and 
    (b) strong scaling, where the total problem size remains fixed as the GPU count increases from 4 to 32. 
    The performance metric reported is the average wall-clock evolve time per time step, selecting the fastest Cartesian domain decomposition ($n_x \times n_y \times n_z = N_{\rm GPU}$) for each configuration. 
    All simulations include the complete physics suite: relativistic electrostatic field solves, synchrotron emission, pair production, and diagnostics. The black dashed line indicates the ideal FFT reference scaling anchored at the average 4-GPU (1 node) baseline across all colliders: $\sim \log N$ for weak scaling and $\sim 1/N$ for strong scaling.}
    \label{fig:scaling_prod}
\end{figure*}

To assess performance, we ran two different weak and strong scaling studies: one with production-like simulations, one with reduced space-charge only simulations. 
In the weak-scaling studies, the number of grid cells and macroparticles is increased in proportion to the number of GPUs, keeping the workload per GPU approximately constant. Specifically, the number of GPUs is doubled at each step and, at the same time, the number of macroparticles and one grid dimension are doubled, cycling through the transverse directions (in order $y$, $x$, $y$, $x$). 
In the strong-scaling studies, the total problem size is fixed while the number of GPUs is increased. The baseline simulations for the scaling studies are case-dependent and are not necessarily the same as the baseline simulations used for the physics comparisons. The physics baselines are selected from the convergence scans, while the scaling baselines are chosen to allow several scaling steps without exceeding GPU memory.

Figure~\ref{fig:scaling_prod} shows the results of the (a) weak-scaling and (b) strong-scaling performed with production-like runs. The performance metric reported is the average wall-clock time per time step.
The timings include the same physics modules and diagnostics used in the benchmark simulations. 
As a reference, the dashed lines show the idealized scaling expected for FFT-only calculations neglecting any other calculations or overheads: $\log N$ for weak scaling and $1/N$ for strong scaling, where $N$ is the number of GPUs and the lines are anchored at the average 4-GPU baseline across all colliders and function only as a guide. 
The largest gaps between WarpX and the ideal FFT-only scaling appear in cases with substantial disruption (ILC) or secondary-particle generation (LWFA), where the extra cost shows up not only in the QED/collision kernels themselves but also in particle redistribution and communication. For the other cases, the departure from ideal scaling appears to be mainly due to FFT-related communication overheads.
Nonetheless, in the weak-scaling study, increasing the problem size by a factor of 8 increases the runtime per step by a factor of 1.4--2.8, with a larger increase for the LWFA case. In the strong-scaling study, increasing the number of GPUs reduces the runtime per step substantially, although the speedup becomes sublinear at larger GPU counts because communication and particle exchange overheads become more important.

The optimal domain decomposition for production runs is not unique and depends on the case and GPU count. 
For each GPU count, all Cartesian MPI decompositions satisfying $n_x n_y n_z = N_{\rm GPU}$ were tested, with $n_u$ the number of MPI ranks along the $u$ axis, and the fastest completed run is reported in Fig.~\ref{fig:scaling_prod}. 
Decompositions with more than one MPI rank along $z$ are often favorable. 
For example, in the FCC-ee~Z weak-scaling scan, the fastest and slowest decompositions on 8 GPUs are $4\times1\times2$ and $8\times1\times1$, respectively, while on 16 GPUs they are $1\times1\times16$ and $4\times1\times4$. 
For the LWFA case, the corresponding decompositions are $4\times1\times2$ and $1\times1\times8$ on 8 GPUs, and $4\times2\times2$ and $1\times2\times8$ on 16 GPUs.
This behavior is not obvious from particle load-balancing considerations alone. 
Since particles move predominantly along $z$, longitudinal decompositions require a systematic handoff of particles between adjacent subdomains. 
In addition, during the collision, subdomains at the smallest and largest $z$ contain fewer particles than the central subdomains. 
One might therefore expect transverse decompositions to be more favorable. 
However, profiler data show that the runtime is dominated by the sliced Poisson solve, through FFT operations, and by inter-rank communications. 
The best decomposition therefore reflects a balance between field-solver cost, communication overhead, and particle load imbalance, rather than particle load balance alone, as often is the case in EM PIC simulations. 
These results show that WarpX extends the accessible simulation regime to larger 3D grids, higher particle counts, and cascade-dominated cases that are impractical with traditional serial beam-beam codes.

\begin{figure}
    \includegraphics[width=\linewidth]{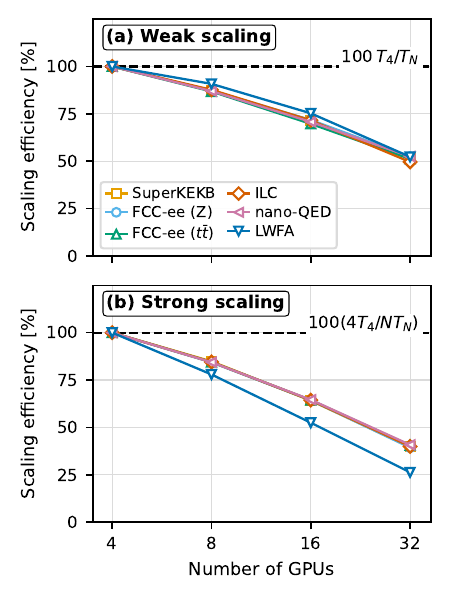}
    \caption{Scaling efficiency of electrostatic WarpX beam-beam simulations without QED physics: (a) weak scaling efficiency, defined as $\eta_{\rm weak} = 100 \times T_4 / T_N$, and (b) strong scaling efficiency, defined as $\eta_{\rm strong} = 100 \times 4 T_4 / (N T_N)$, where $T_N$ is the average wall-clock time per step on $N$ GPUs. The horizontal dashed line at 100\% denotes ideal scaling efficiency. The domain decomposition is fixed to $1\times1\times N$, where $N$ is the GPU count.}
    \label{fig:scaling_space_charge}
\end{figure} 

To isolate the performance of the core numerical field solver from other case-specific effects, Fig.~\ref{fig:scaling_space_charge} shows the results of the (a) weak-scaling and (b) strong-scaling studies with reduced simulations. Here, we disabled all QED physics and only kept the space-charge calculations, i.e. the Poisson solve. The optimal decomposition in this simplified simulations is always $1\times1\times N_{\rm GPU}$. The metric we use in this case is the scaling efficiency, defined as 
\begin{equation}
\eta = \begin{cases}
    100 \times T_4 / T_N & \rm weak \\
    \quad \\
    100 \times 4 T_4 / (N T_N) & \rm strong,
\end{cases}
\end{equation}
where $T_N$ is the average time per timestep on $N$ GPUs. 
An ideal efficiency of 100\% means that every worker carries the same workload regardless of the problem size in the weak scaling, and that the fixed workload is divided equally regardless of the number of workers in the strong scaling. This would be only possible if communications between GPUs were not needed and overheads were not present. In contrast to the production runs, all six collider cases collapse onto a universal scaling curve. In weak scaling, the solver achieves approximately 90\% efficiency on 8 GPUs, 72\% on 16 GPUs, and 50\% on 32 GPUs. This degradation is driven by the inter-node communication overhead of the distributed 2D FFT transposes required for open boundary conditions, particle redistributions across GPUs, as well as the intrinsic $\mathcal{O}(N \log N)$ complexity of the algorithm. In strong scaling, the solver maintains over 65\% efficiency up to 16 GPUs and $\sim40\%$ on 32 GPUs for the circular collider configurations, providing a $\times 3.2$ speedup over the 4-GPU baseline. The LWFA case exhibits lower strong-scaling efficiency at 32 GPUs ($\sim 26\%$) because of its smaller baseline size. This results in local subdomains that under-utilize GPU resources once partitioned across eight nodes. The baseline problem size for LWFA was deliberately chosen to be smaller to prevent secondary particle generation from exhausting GPU memory during production runs. 

Taken together, these results confirm that WarpX effectively balances distributed spectral field solves with intense particle workloads, providing the parallel scalability required for large-scale, self-consistent 3D beam-beam simulations across a broad range of collider scenarios.

\section{\label{sec:conclusions}Conclusions}
In this paper, we have presented WarpX as a GPU-accelerated particle-in-cell code for beam-beam simulations at the interaction point of different lepton-based colliders. 
We benchmarked WarpX against the established beam-beam codes GUINEA-PIG and CAIN for a broad range of collider-relevant scenarios, including SuperKEKB, FCC-ee, ILC, and laser-wakefield-accelerator beams. For the beam and numerical parameters considered here, WarpX reproduces the main beam-beam observables obtained with these codes, including luminosity, photon, and pair-production distributions.

These benchmarks demonstrate that WarpX captures the physics modeled by traditional beam-beam tools, while extending the accessible simulation regime through a multi-GPU parallel PIC framework. This GPU-distributed parallelization makes it possible to use larger 3D grids, higher macroparticle numbers, and sustain larger secondary-particle populations within practical wall-clock times. This capability is especially important for scenarios in which high-resolution modeling, strong disruption, or secondary-particle cascades become computationally demanding. While the present work focuses on $e^+e^-$ collider applications, the same numerical framework can be applied more broadly. We plan to use WarpX to study beam-beam interactions in the Electron-Ion Collider (EIC) and Muon Collider (MC). 

WarpX provides an open-source, actively developed, and portable platform for high-fidelity beam-beam simulations on modern computing architectures. By combining self-consistent multi-physics modeling and multi-GPU acceleration in a single workflow, WarpX enables large-scale beam-beam simulations that were not feasible before and, ultimately, provides a new capability for the collider community that can serve as a foundation for next-generation beam-beam modeling.

\begin{acknowledgments}
The authors acknowledge for fruitful discussions during the course of this research the following people: Simone Pagan Griso, Simon Knapen, Angira Rastogi, Christiane Scherb, Inbar Savoray (LBNL), Michael Peskin, Caterina Vernieri, Dimitris Ntounis, Laith Gordon (SLAC), Jan Eysermans (CERN), Katie Kudela (MIT), and Luca Fedeli (CEA).
We extend our special thanks to the high-school students who participated in the Experiences in Research program at LBNL in 2024 and 2026: Alisa Garduno, Ayush Bansal, Dionne Wong, Evaline Flamer, Leo Sitaraman, Nicholas Yan.  
This material is based upon work supported by the CAMPA collaboration, a project of the U.S. Department of Energy, Office of Science, Office of Advanced Scientific Computing Research and Office of High Energy Physics, Scientific Discovery through Advanced Computing (SciDAC) program.
This work was also supported by the Laboratory Directed Research and Development Program of Lawrence Berkeley National Laboratory under U.S. Department of Energy Contract No. DE-AC02-05CH11231.
This research used resources of the National Energy Research Scientific Computing Center (NERSC), a Department of Energy Office of Science User Facility using NERSC award HEP-ERCAP0031191.
\end{acknowledgments}

\appendix

\section{Algorithms in GUINEA-PIG}
\label{sec:app_gpalgo}

We describe the processes enabled in the benchmarks using GUINEA-PIG (C++ version 1.2.2~\cite{guineapig}): the collective fields, beamstrahlung, Compton scattering, radiative Bhabha scattering, and coherent and incoherent pair production. Other GUINEA-PIG processes were disabled.

GUINEA-PIG represents bunch $b\in\{1,2\}$ by $N_{m,b}$ macroparticles of weight $w_b=N_b/N_{m,b}$ and divides each bunch into $N_s$ longitudinal slices. During the $2N_s-1$ longitudinal steps, slices from opposing beams interact when $s_1+s_2=k$. Each step contains $N_t$ transverse substeps. At every substep, the code deposits charge on a 2D transverse grid, recomputes the collective fields, advances charged particles in
the opposing field, and evaluates QED processes. Longitudinal motion is represented by the changing slice overlaps rather than by explicitly advancing particles through a fixed longitudinal mesh.

Algorithm~\ref{alg:collision} summarizes the collision loop. 
Process-specific sampling details are described below.

\begin{lstlisting}[breaklines=true,breakatwhitespace=true,
                   caption={beam-beam collision loop in GUINEA-PIG.},
                   label={alg:collision}]
initialize bunches and longitudinal slices
for longitudinal step $k=0,\ldots,2N_s-2$ do
  identify overlapping slices with $s_1+s_2=k$
  for each transverse substep do
    for each overlapping slice pair do
      deposit primary and coherent-pair charge
      calculate cell-based luminosity
      solve for the two transverse potentials
      advance primary particles and coherent pairs
        in the opposing field
      sample beamstrahlung emission
      sample cell-based Compton scattering and
        incoherent pair production
      transport beamstrahlung photons and sample
        coherent pair conversion
      transport tracked secondary charged particles
    end for
  end for
end for
\end{lstlisting}

Beamstrahlung is sampled as quantum synchrotron radiation experienced by each charged macroparticle (electron or positron) in the local EM field. 
In a substep, the longitudinal and angular displacements of the macroparticle are $\Delta \ell = \Delta z / (2 N_t)$, with $\Delta z$ the slice length, and $\Delta \theta = |\Delta \textbf{v}_\perp|/c$. Here, $\Delta \textbf{v}_\perp$ is the change in transverse velocity during the substep and is calculated from the collective field of the opposing beam $\Delta \textbf{v}_\perp =  \pm 2e\textbf{E}_\perp \Delta \ell/ (\gamma  m_e c ) $, where $\gamma$ is the Lorentz factor of the macroparticle at the beginning of the substep. Then, the local bending radius of the trajectory is $R=\Delta \ell / \Delta \theta$.  
The beamstrahlung parameter of the macroparticle is then $\Upsilon=\bar\lambda_c \gamma^2/R$, with $\bar \lambda_c$ the reduced Compton wavelength.
GUINEA-PIG samples photon emission and energy by rejection sampling~\cite{williamwxgp} from the local Sokolov-Ternov quantum synchrotron spectrum~\cite{sokolov1968synchrotron}:
\begin{equation}
\frac{d^2N}{dy dt}  = \frac{\alpha c}{\sqrt{3} \pi \bar{\lambda}_c \gamma} \left[  \int_\xi^\infty \! \! K_{5/3}(\xi') d \xi' + \frac{y^2}{1-y} K_{2/3}(\xi) \right], 
\label{eq:sokter}
\end{equation}
where $\gamma_\gamma=\hbar\omega/(m_ec^2)$ is the photon energy in units of $m_e c^2$, $y=\gamma_\gamma/\gamma$ is the fractional photon energy, $\alpha$ the fine-structure constant,
$$
\xi = \frac{2}{3\Upsilon} \frac{y}{1-y},
$$
and $K_\nu(s) $ is the modified Bessel function of the second kind of order $\nu$.
$d^2N$ can be interpreted as the expected number of photons emitted by a single electron/positron with normalized energy in an interval $dy$ around $y$, in a time interval $dt$. 
GUINEA-PIG divides the substep into $N_{rad} = \lfloor 10 P_0 \rfloor + 1$ radiation trials, where $P_0 = 4^{1/3} \alpha \gamma \Delta \theta / \Gamma(4/3) $ is an upper estimate of the number of emitted photons. Each trial is assigned $\delta \theta = \Delta \theta / N_{rad}$. A trial proposes a photon if a uniform random number is smaller than
$$
p_0 = \frac{\alpha C_0}{\sqrt{3} \pi }\frac{\gamma \delta \theta}{(1 + 3\Upsilon/4)^{1/3} }
$$
with $C_0 = 9 \times 2^{-1/3} \Gamma(2/3)$. If a photon is proposed, its energy $y$ is drawn from 
$$
y = \frac{3\Upsilon u^3/2}{1 - u^3  + 3\Upsilon ( 1+u^6)/4}
$$
where $u\sim \mathcal{U}(0,1)$ is a random variable uniformly distributed in $(0,1)$. 
This proposed value is kept with probability 
$$
g = F(y) \frac{dy}{du} \frac{(1+3\Upsilon/4)^{1/3}}{C_0 \Upsilon}, 
$$
where $F(y)$ is an approximation of the Sokolov-Ternov bracket in Eq.~\ref{eq:sokter} that uses piecewise fits to evaluate the Bessel functions therein.
The photon energy $\gamma_\gamma = y \gamma $ is subtracted from the parent particle, whose energy becomes $\gamma ( 1-y)$. Each later trial rescales $\delta \theta$ to $\delta \theta / (1-y)$, to keep $p_0$ constant. $\Upsilon$ is also fixed throughout the trials and a later photon in the same substep carries a fraction of the energy left after the earlier ones. Photons are created at the parent particle's end-of-substep position and travel parallel to it, without explicit transverse recoil. This algorithm is taken from CAIN~\cite{Yokoya:1985ab}.

A real beamstrahlung photon can convert into a coherent $e^+e^-$ pair in the opposing bunch field. 
GUINEA-PIG evaluates coherent photon conversion in the local opposing-bunch field using an approximate strong-field pair-creation rate~\cite{erber_rmp_1966}.
Analogous to Eq.~\ref{eq:sokter}, the spectral rate of photon decays into electron-positron pairs is~\cite{sokolov1968synchrotron}: 
\begin{align}
\frac{d^2N_\pm}{dx dt} = & \frac{\alpha c}{ \sqrt{3} \pi \bar \lambda_c \gamma_\gamma }
\left[  \int_\eta^\infty K_{1/3} (\eta') d\eta'   \right. \nonumber \\ 
& + \left(  \frac{x}{1-x} \left. + \frac{1-x}{x} \right) K_{2/3} (\eta)  \right], 
\label{eq:sokter+-}
\end{align}
where $x =\gamma_+ / \gamma_\gamma$ is the fractional energy of the positron and
$$
\eta = \frac{2}{3 \Upsilon_\gamma x(1-x)}.
$$
$d^2N_\pm$ can be interpreted as the expected number of positrons with fractional energy within a neighborhood $dx$ around $x$, in a time interval $dt$. 
The code calculates the local SF-QED photon parameter of a macrophoton as $\Upsilon_\gamma= \gamma_\gamma |2 \mathbf E_\perp|/ E_{\rm cr}$, where the field is evaluated at its position at the beginning of the substep. 
The photon is then advanced by $\Delta \ell = \Delta z / (2 N_t)$.
The code approximates the integrated pair-conversion probability over the substep as~\cite{erber_rmp_1966}:
\begin{equation}
    P = \frac{\alpha^2 \Delta \ell }{r_e \gamma_\gamma} \left[0.23 \Upsilon_\gamma  \exp \! \left( - \frac{8}{3 \Upsilon_\gamma}\right) (1+0.22 \Upsilon_\gamma)^{-1/3} \right]. 
\end{equation}
The code draws a uniform random number $u$.
If $u \geq 1.36 P$, then the photon is not converted. 
If $u < 1.36 P $ and $1.36 P < 0.1$, there is a conversion attempt. 
If $u < 1.36 P$ and $1.36 P \geq 0.1$, the attempt is repeated up to $N = \lfloor 1.36 P / 0.1 \rfloor + 1 $ times, and each trial is kept with probability $1.36 P / N$, stopping at the first success.
An attempt samples the positron energy fraction $x$ from the spectrum in Eq.~\ref{eq:sokter+-}, using a spline table for the Bessel functions.
On conversion, i.e. when the first $x$ is accepted, two macroparticles (one electron and one positron) are created with energies $x \gamma_\gamma$ and $(1-x) \gamma_\gamma$. Both particles inherit the photon's post-step position and transverse direction, while the source photon is removed. Coherent pairs remain attached to their parent slice. They are transported, can radiate, and deposit charge in subsequent substeps, so they can affect later collective-field solves. 

Incoherent reactions are evaluated between macroparticles from opposing beams in the same transverse cell. Real photons ($\gamma$) are beamstrahlung photons. Virtual photons ($\gamma^\ast$) are sampled from primary beam particles using the Weizs\"acker-Williams approximation.
The three incoherent pair channels are Breit-Wheeler ($\gamma\gamma\to e^+e^-$), Bethe-Heitler ($\gamma\gamma^\ast\to e^+e^-$), and Landau-Lifshitz ($\gamma^\ast\gamma^\ast\to e^+e^-$). GUINEA-PIG uses the (linear) Breit-Wheeler cross section and angular distribution at the sampled photon energies for all three. Virtual-photon spectra and the configured $Q^2$ condition account for virtuality; virtuality is otherwise neglected in the pair kinematics. The benchmark setting \texttt{pair\_q2=2} requires $Q_a^2,Q_b^2\leq s_{\gamma\gamma}/4$.

The same cell-based calculation handles photon-lepton scattering. A real photon represents ordinary Compton scattering; a virtual photon represents radiative Bhabha scattering (bremsstrahlung). Both use the Klein-Nishina total and differential cross sections, subject to the configured virtuality and energy cuts. The optional beam-size displacement of virtual photons was disabled in these benchmarks (\texttt{beam\_size=0}).

For a candidate collision with statistical weights $W_a$ and $W_b$, GUINEA-PIG sets the expected event count to $\mu=W_aW_b\sigma$. It performs $N=\lfloor\mu\rfloor+1$ trials, each accepted with probability $\mu/N$. Accepted incoherent pairs and Compton-scattered charged particles can be tracked through the collective fields but are not deposited on the charge mesh. Outgoing Compton photons can be written to output but are not transported as secondaries. These incoherent calculations do not remove or modify the incoming beam macroparticles or photons.

\section{Algorithms in WarpX}
\label{sec:app_wxalgo}

We describe the processes enabled in WarpX (version 26.09~\cite{warpx2609}) benchmarks: the collective (electrostatic) beam-beam interaction, beamstrahlung, coherent pair creation where applicable, Compton scattering, radiative Bhabha scattering, and the three incoherent pair-production channels in Table~\ref{tab:bbprocesses}.

WarpX represents each bunch by macroparticles of initial weight $w_b=N_b/N_{m,b}$ and evolves both bunches on a shared 3D mesh. The benchmark time step is $\Delta t=\Delta z/(2c)$: two counterpropagating, nearly light-speed beams therefore move approximately one longitudinal cell relative to one another per step.
Unlike GUINEA-PIG's overlapping bunch slices, WarpX's longitudinal mesh planes are fixed, and particles and photons move through them. Third-order particle shapes are used for deposition and field gathering.

The collective field model is relativistic electrostatic PIC. For each depositing charged species $s$, WarpX deposits $\rho_s$ and calculates one mean velocity $\overline{\boldsymbol\beta}_s c$ for the species. The benchmarks use an integrated Green-function solver that solves independently on each transverse plane~\cite{quiang_igf2d}:
\begin{equation}
 \nabla_\perp^2\phi_s(x,y,z_k)
 =-\frac{\rho_s(x,y,z_k)}{\epsilon_0}.
 \label{eq:wx_slice_poisson}
\end{equation}
The species fields are reconstructed from their potentials and summed before being gathered by charged particles. This solve neglects longitudinal coupling between planes in the potential, although the deposited charge varies with $z$ and field reconstruction uses derivatives in all three directions. A full 3D solver is optionally available~\cite{IGFsolver3d}, which would allow investigating the longitudinal coupling.
Using a single average velocity does not exactly resolve velocity spread within a species, and the model does not propagate electromagnetic waves. Charged particles are advanced with the Vay pusher~\cite{JLV2008}.

Algorithm~\ref{alg:wx_collision} summarizes a time step. 
It omits the momentum half-step placement of collision calls.

\begin{lstlisting}[breaklines=true,breakatwhitespace=true,
                   caption={beam-beam collision loop in WarpX.},
                   label={alg:wx_collision}]
initialize bunches and collective fields
for each physical time step do
  sample enabled beamstrahlung and coherent pair conversion
    using the current fields
  regenerate virtual photons from configured primary leptons
  sample configured photon-photon and photon-lepton
    reactions within 3D mesh cells
  advance charged particles and transport real photons
  for each charged species that deposits charge do
    deposit charge and calculate its mean velocity
    solve the transverse Poisson equation on each $z$ plane
    reconstruct its electric and magnetic fields
  end for
  sum species fields for the next step
end for
\end{lstlisting}

Primary beam particles deposit charge. Coherent pairs also deposit charge and can affect subsequent field solves. Incoherent pairs and Compton-scattered charged particles are transported through the collective fields but do not deposit charge.

Beamstrahlung and coherent pair creation use the locally constant field approximation implemented in PICSAR-QED~\cite{fedeli2022picsar}. For a lepton, the local strong-field parameter is
\begin{equation}
 \chi_\ell=\frac{\gamma_\ell}{E_{\rm cr}}
 \sqrt{(\boldsymbol E+\boldsymbol v_\ell\times\boldsymbol B)^2
       -(\boldsymbol v_\ell\cdot\boldsymbol E)^2/c^2}.
 \label{eq:wx_chi_lepton}
\end{equation}
Tabulated rates reduce particle or photon optical depths until an event occurs; lookup tables then sample the emitted photon energy or the pair-energy split. 
Emitted photons are collinear with their parent lepton, and their momentum is subtracted from it. 
On coherent conversion, the source photon is removed and a nearly collinear electron-positron pair is created. Classical continuous radiation reaction is disabled, and the configured $\chi$ threshold suppresses emission below the tabulated model's lower limit.

Incoherent reactions use cell-based binary collisions~\cite{higginson2020corrected}.
At the start of each PIC time step, before the collision modules run, WarpX regenerates virtual photons from the configured primary leptons using the angle-integrated Weizs\"acker-Williams spectrum~\cite{LandauVol4},
\begin{equation}
 \frac{dN_{\gamma^\ast}}{dy}
 \simeq\frac{2\alpha}{\pi}\frac{\ln(1/y)}{y},
 \qquad y_{\min}\le y\le 1,
 \label{eq:wx_virtual_spectrum}
\end{equation}
where $y=\mathcal{E}_{\gamma^\ast}/\mathcal{E}_p$ and $y_{\min}=\mathcal{E}_{\min}/\mathcal{E}_p$.
A sampling factor multiplies the number of virtual-photon macroparticles and divides each of their weights by the same factor, so Eq.~\eqref{eq:wx_virtual_spectrum} is the physical spectrum.
Virtual photons are emitted along the parent momentum and placed at the parent position.
The optional beam-size displacement is available in 3D and is disabled in our simulations.
When it is enabled, the code samples a virtuality $Q^2$ and displaces the photon perpendicular to the parent momentum by $\rho=\hbar/\sqrt{Q^2(1-y)}$~\cite{PhysRevAccelBeams.27.091001}.
The virtual-photon population is generated once per PIC step and is reused if a collision module is subcycled within that step.

Macroparticles of the two configured species that share a cell are shuffled and paired at random.
Pairing does not depend on direction; the pair kinematics sets the cross section.
For two photon species the reaction is linear Breit-Wheeler pair production.
The channel is selected by which species enter the pair, within the equivalent-photon approximation: real-real, real-virtual, and virtual-virtual collisions correspond to the Breit-Wheeler, Bethe-Heitler, and Landau-Lifshitz processes.
A real or virtual photon paired with a lepton uses the Klein-Nishina total cross section for the reaction probability and the Klein-Nishina differential cross section for the scattering angle, corresponding to Compton scattering or radiative Bhabha scattering.

The probability of each sampled pair is weighted so that the combinations omitted by the pairing are included~\cite{higginson2020corrected,remipasc}.
For two distinct species, $N_{\rm partners}=\min(N_1,N_2)$, where $N_i$ is the number of macroparticles of species $i$ in the cell, and macroparticles of the smaller group are reused.
With $w_{\rm max}=\max(w_1,w_2)$, the linear Breit-Wheeler probability of one sampled pair is
\begin{equation}
P=1-\exp\!\left(
-N_{\rm partners}\,M\,
\frac{E_1^\star E_2^\star}{E_1 E_2}\,
\sigma\,v_\star\,w_{\rm max}\,
\frac{\Delta t}{\Delta V}
\right),
\label{eq:wx_pair_probability}
\end{equation}
where $\sigma$ and the relative velocity $v_\star$ are evaluated in the center-of-momentum frame ($v_\star=2c$ for two photons), and $E_i$ and $E_i^\star$ are the lab-frame and center-of-momentum energies.
Compton scattering uses the same pairing and the same weights.
Its rate is evaluated in the lepton rest frame, with $v=c$, $\sigma$ the Klein-Nishina total cross section, and the energy ratio in Eq.~\eqref{eq:wx_pair_probability} replaced by $E_{\gamma,\mathrm{rest}}/(E_{\gamma,\mathrm{lab}}\gamma_\ell)$.
The event multiplier $M\ge 1$ turns rare reactions into more product macroparticles of smaller weight.
When the trial probability of a pair exceeds the configured threshold, that pair uses a reduced multiplier, floored at one.
That reduced value is the $M$ in Eq.~\eqref{eq:wx_pair_probability} and in the product weight. 
When a reaction is accepted, the product macroparticles are created with weight $w_{\rm min}/M$, where $w_{\rm min}=\min(w_1,w_2)$, and that weight is subtracted from both colliding macroparticles, including a virtual photon.
The primary lepton keeps its weight, because a virtual photon stores no parent index.
Regeneration at the next time step replaces the virtual photons, so this debit remains on the virtual-photon population of the current step.
A colliding macroparticle whose weight reaches zero is removed for the rest of the step.
Real product photons are added to their configured species, transported with that species, and can undergo further collisions.

\section{Approximate head-on boosted frame}
\label{sec:app_boost}
\begin{figure*}
    \includegraphics[width=\linewidth]{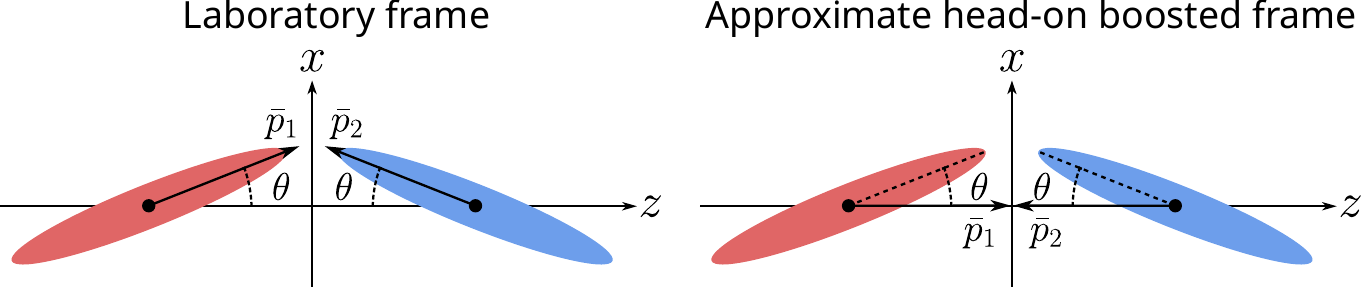}
    \caption{beam-beam collision in two frames of reference. Left: laboratory frame, where the beams collide at an angle. Right: approximate boosted frame where the beam particles are rotated in space and collide head-on, without any space-time mixing.}
    \label{fig:approx_boost}
\end{figure*}

Consider two ultrarelativistic beams that collide with a half crossing angle $\theta$ in the $z$--$x$ plane of the laboratory frame, as shown in Fig.~\ref{fig:approx_boost}. In this section, we use $c=1$. We distinguish the nominal beam momenta $\bar{p}_i^\mu = (\bar{E}_i,\, \bar{\textbf{p}}_i) $ from the four-momentum $p_i^\mu = (E_i,\, \textbf{p}_i)$ of an arbitrary beam particle, with $i=1,2$. 
The nominal three-momenta are
\begin{align*}
 \bar{p}_{ix} = \bar p_i \sin \theta, \,  
 \bar{p}_{iy} = 0, \, 
 \bar{p}_{iz} = \pm \bar p_i \cos \theta
\end{align*}
where $\bar p_i  = |\bar{\textbf{p}}_i| $.

A Lorentz boost purely along $x$ 
\begin{equation}
    \Lambda_{\text{boost}} = 
    \begin{pmatrix}
        \gamma & -\gamma \beta & 0 & 0 \\
        -\gamma \beta & \gamma & 0 & 0 \\
        0 & 0 & 1 & 0 \\
        0 & 0 & 0 & 1 
    \end{pmatrix}
\end{equation}
transforms $\bar{p}_i^\mu$ into $ (\bar{p}'_i)^\mu = \Lambda_{\text{boost}} \bar{p}_i^\mu$ with
\begin{align*}
\bar p'_{ix} = \gamma (\bar p_{ix} - \beta \bar E_i), \quad \bar p'_{iz} = \bar p_{iz}.
\end{align*}
If both beams are ultrarelativistic,
\begin{equation}
\frac{\bar p_{ix}}{\bar E_i}
=
\frac{\bar p_i}{\bar E_i}\sin\theta
\simeq \sin\theta,
\qquad i=1,2.
\end{equation}
We therefore introduce a common transverse boost 
\begin{equation}
\beta_{\mathrm{boost}}=\sin\theta,
\qquad
\gamma_{\mathrm{boost}}=\sec\theta = 1/\cos\theta, 
\end{equation}
whose transformation matrix is
\begin{equation}
    \Lambda_{\text{boost}} = 
    \begin{pmatrix}
          \sec \theta & - \tan \theta & 0 & 0 \\
        - \tan \theta & \sec \theta  & 0 & 0 \\
        0 & 0 & 1 & 0 \\
        0 & 0 & 0 & 1
    \end{pmatrix}. 
\end{equation}

Upon this transformation, the nominal momenta become approximately antiparallel, up to ultrarelativistic corrections, according to
\begin{align}
\bar p'_{ix} & = \sec\theta\,\bar p_i\sin\theta -\tan\theta\,\bar E_i \nonumber \\
& = \tan\theta\,(\bar p_i-\bar E_i) \simeq 0, \label{eq:ppix}
\end{align}
where the last approximation follows from 
$$
\bar p_i - \bar E_i = - \frac{m_i}{2  \Gamma_i } + \mathcal{O}(m_i \Gamma_i^{-3}) , 
$$
where $\Gamma_i\gg 1$ is the beam-particle nominal Lorentz factor.
For small crossing angles, the nominal energy and momentum coincide with their head-on values through first order in $\theta$: 
\begin{align*}
\bar E'_i = \bar E_i+\mathcal O(\theta^2), \quad \bar p'_{iz} &= \pm\bar p_i+\mathcal O(\theta^2).
\end{align*}
Upon the considered boost, the four-position of an arbitrary beam particle transforms as
\begin{align}
 {x'_i}^\mu & = \Lambda_{boost} x_i^\mu \nonumber \\
& = \left( \sec \theta \, t_i - \tan \theta \, x_{i}, \,
           \sec \theta  \, x_i - \tan \theta \, t_i , \,
           y_i, \,
           z_i \right) \nonumber \\ 
& = \left( t_i - \theta x_i, \, 
                 - \theta t_i + x_i, \, 
                  y_i, \, 
                  z_i \right)  +\mathcal O(\theta^2) \nonumber \\
& = \left( t_i, \, x_i, \, y_i, \, z_i \right) +\mathcal O(\theta) \label{eq:xboost}.
\end{align}
For particles initialized at the common laboratory time $t_i=0$, the spatial coordinates are unchanged by the boost through first order in $\theta$, while the transformed time contains a first-order offset.
The four-momentum transforms as
\begin{align*}
{p'_i}^\mu & = \Lambda_{boost} p_i^\mu \nonumber \\
 & = \left( \sec \theta \, E_i - \tan\theta \, p_{ix} , \sec \theta \, p_{ix} - \tan\theta \, E_i, p_{iy}, p_{iz} \right) \nonumber \\
& = \left( E_i - \theta p_{ix}, \, p_{ix} - \theta E_i  , \, p_{iy}, \, p_{iz}\right) + \mathcal{O}(\theta^2) .
\end{align*}

In summary, for ultrarelativistic beams, the transverse boost transforms the nominal momenta into their head-on values through first order in $\theta$ and up to corrections in the ultrarelativistic approximation.
For the particles initialized at the laboratory time $0$, the spatial coordinates remain unchanged to first order, while the boost introduces a position-dependent time offset $t'_i = - \theta x_i + \mathcal{O}(\theta^2)$. The approximate boosted-frame representation used in GUINEA-PIG and WarpX retains the first-order transformation of the particle momenta while retaining only the zeroth-order transformation of the four-positions. It therefore uses the rotated laboratory-frame particle positions together with the corresponding head-on momenta, neglecting the space-time mixing. Note that WarpX supports simulations in exact boosted frame, which will be considered in future work.

\bibliography{ref}

\end{document}